\documentclass[aip,jcp,reprint]{revtex4-1}
\usepackage{colordvi,epsfig,color,amsmath,amssymb,bm}

\def\be{\begin{equation}} 
\def\ee{\end{equation}} 
\def\bee{\begin{eqnarray}} 
\def\eee{\end{eqnarray}}

\def\kb{k_{\rm B}}
\def\tilde{\widetilde}
\def\halb{\mbox{$\frac{1}{2}$}}

\newcommand{\bbbone}{{\mathchoice {\rm 1\mskip -4mu l}{\rm 1\mskip -4mu l}{\rm
1\mskip -4.5mu l}{\rm 1\mskip -5mu l}}}

\definecolor{darkgreen}{rgb}{0,0.5,0}
\definecolor{purple}{rgb}{0.35,0,0.35}
\definecolor{orange}{rgb}{1,0.5,0}
\definecolor{darkred}{rgb}{.7,0,0}
\definecolor{darkblue}{rgb}{0,0,.3}
\definecolor{grey}{rgb}{.6,.6,.6}
\definecolor{dimgreen}{rgb}{0.2,0.6,0.1}

\begin{document}

\title{Quasi-adiabatic path integral approach for quantum systems under the influence of multiple non-commuting fluctuations}

\author{T. Palm$^{1,2}$ and P. Nalbach$^{1}$}
\affiliation{
$^1$Fachbereich Wirtschaft \& Informationstechnik, Westf\"alische Hochschule, M\"unsterstrasse 265, 46397 Bocholt, Germany\\
$^2$I.\ Institut f\"ur Theoretische Physik, Universit\"at Hamburg,
Jungiusstra{\ss}e 9, 20355 Hamburg, Germany
}

\date{\today}

\begin{abstract}

Quantum systems are typically subject to various environmental noise sources. Treating these environmental disturbances with a system-bath approach beyond weak coupling one must refer to numerical methods as, for example, the numerically exact quasi-adiabatic path integral approach. This approach, however, cannot treat baths which couple to the system via operators, which do not commute. We extend the quasi-adiabatic path integral approach by determining the time discrete influence functional for such non-commuting fluctuations and by modifying the propagation scheme accordingly. We test the extended quasi-adiabatic path integral approach by determining the time evolution of a quantum two-level system coupled to two independent bath via non-commuting operators. We show that convergent results can be obtained and agreement with analytical weak coupling results is achieved in the respective limits.

\end{abstract}


\maketitle

\section{Introduction}

Open quantum dynamics is a very successful approach to describe and treat dissipative effects like relaxation, decoherence and dephasing in quantum systems \cite{Caldeira1981, Leggett1987, WeissBuch, BreuerRMP2016}. Dissipation results therein by coupling the quantum system of interest to an environment. 
The later is typically described by a set of harmonic oscilators bilinear coupled to the system. The according system-bath model can then be treated either perturbatively or by numerical exact methods. Thus, successful treatment of, for example, problems like energy transfer in photosynthetic complexes \cite{PhotosyntheticExcitons-book-2000, May-Kuehn-2011, NalbachNJP2010} and dephasing in various qubit realizations \cite{SchoenRMP2001, NalbachLZ2013} is enabled.

Typically, the quantum systems of interest are subject to various noise sources. Charge and flux qubits, for example, experience noise due to phonons, voltage fluctuations in the various gates, charged defects and currents through nearby quantum point contacts \cite{SchoenRMP2001, NalbachLZ2013, RPP2006}. Chromophores in photosynthetic complexes are disturbed by strong environmental fluctuations due to intra- and intermolecular vibrations of the photoactive complexes, vibrations of embedding proteins, solvent fluctuations and the charge separation in the reaction center \cite{May-Kuehn-2011}. 

For multiple bath cases, typically, simplifying approaches are employed.
If one noise source dominates, the others are usually neglected. If all environmental influences are weak, a weak coupling treatment for each noise source can be done and the effects are  {\it added independently}. Another option is to treat one bath phenomenologically (by introducing a rate) and only treating a second bath within a system-bath approach \cite{Petta2010b}. Standard methods can easily be extended to this problem. This approach fails to describe the dynamics correctly when the phenomenological treated noise is strongly coupled to the system \cite{PalmPRA2017} even at very weak coupling to the other noise sources.

In case that multiple noise sources are present and all are strongly coupled to the quantum system of interest all these approaches fail and one must treat all baths on equal footing. To this end, we extend here the numerical exact quasi-adiabatic path integral approach (QUAPI) \cite{MakriQUAPI1995a, MakriQUAPI1995b} to treat multiple environments. For baths which couple to the system via operators, which commute, an according extension of QUAPI is straight forward \cite{NalbachNJP2010}. If, however, these coupling operators do not commute, the resulting non-commuting fluctuations in the system can give rise to peculiar non-perturbative effects like quantum frustration of decoherence \cite{CastroPRL2003, KehreinPRB2013}. To treat this case with QUAPI, we determine the necessary time discrete form of the total influence functional for a generic case of a quantum system subject to two noise sources and then extend QUAPI accordingly. 

We demonstrate in section IV the functionality of our extended QUAPI by calculating the dynamics of a quantum two-level system (TLS) coupled to two independent bath each coupling to a separate Pauli matrix. We compare our results with various single bath cases. At weak coupling our results coincide with standard analytical approximative results. At first, however, in the next section we introduce the model, the time-discretized time evolution of the reduced density of states and the necessary influence functionals. In the third section we present the extended QUAPI scheme for a case with two independent bath. We end with a conclusion.

\section{The dissipative influence functional}

Throughout this paper we deal with a generic system-bath Hamiltonian of the form
\be
H = H_S + H_{SB,1} + H_{SB,2} .
\ee
The system with Hamiltonian $H_S$ is disturbed by two independent fluctuation sources modelled as harmonic baths
\be\label{eq2.3}
H_{SB,\nu} = \sum_{k=1}^M \frac{p_{k,\nu}^2}{2m_{k,\nu}} + \halb m_{k,\nu} \omega_{k,\nu}^2 \left( q_{k,\nu} - \frac{\lambda_{k,\nu} \hat{\sigma}_{\nu}}{m_{k,\nu}\omega_{k,\nu}^2} \right)^2 
\ee
and $[q_{k',\nu'},p_{k,\nu}]=i\hbar \delta_{k,k'}\delta_{\nu,\nu'}$. Herein, the $q_{k,\nu}$ and $p_{k,\nu}$ are the position and momentum of mode $k$ with frequency $\omega_{k,\nu}$ of bath $\nu$ coupled via $\lambda_{k,\nu}$ to the system. The $\hat{\sigma}_\nu$ are system operators, i.e. the system part of the system-bath coupling, with eigenvectors $|\sigma_\nu\rangle$ and corresponding eigenvalue $\sigma_\nu$. All relevant information about the baths are captured in their spectral densities
\be\label{eq2.4}
G_\nu(\omega)=\sum_{k=1}^M \frac{\lambda_{k,\nu}^2}{2m_{k,\nu} \omega_{k,\nu}} \delta(\omega-\omega_{k,\nu}) .
\ee

All equations are given for a general system Hamiltonian with a countable set of eigenstates. 

\subsection{Abelian Fluctuations}

We seek the time dependent reduced density matris of the system $\rho_{S,r}(t)={\rm Tr}_B\{ \rho(t) \}$ achieved by averaging out the baths degrees of freedom from the statistical operator $\rho(t)$ of system plus bath. For simplicity, we assume a factorized initial condition for the total density matrix $\rho(t)$, i.e. 
\be 
\rho(0)=\rho_S(0)\rho_B(0)
\ee
with $\rho_{S}(0)$ and $\rho_{B}(0)$ the initial statistical operator of the system and bath respectively.

If the states $|\sigma\rangle$ are eigenvectors with eigenvalues $\sigma$ to the system-bath coupling operators $\hat{\sigma}_1$ and $\hat{\sigma}_2$, a representation for the reduced density matrix can be given in terms of a path integral whereby the baths influence is captured within a Feynman - Vernon influence functional \cite{Fey63} $I(\cdot)$, i.e.
\bee
&& \rho_{S,r}(\sigma' , \sigma'' ;t) = {\rm Tr}_B\left\{ \langle \sigma'' | e^{-iHt/\hbar} \rho(0) e^{iHt/\hbar} | \sigma' \rangle\right\} \label{rhored}\\
&&= \prod_{j=0}^{N-1} \int d\sigma_j^+ \int d\sigma_j^- \langle \sigma_{j+1}^+ | e^{-iH_S \delta t/\hbar} |\sigma_j^+\rangle   \langle \sigma_{0}^+ | \rho_S(0) |\sigma_0^-\rangle \nonumber \\
&& \hspace*{1cm}\times \langle \sigma_{j}^- | e^{iH_S \delta t/\hbar} |\sigma_{j+1}^- \rangle \cdot I(\sigma_0^+, \ldots,\sigma_{N}^+,\sigma_0^-,\ldots,\sigma_N^-) \nonumber
\eee
with $\sigma'=\sigma_N^-$ and $\sigma''=\sigma_N^+$. Herein, we explicitly employed a Trotter time slicing with $N$ slices of duration $\delta t$ which is advantegeous for numerical evaluation. Makri and Makarov developed the quasi-adiabatic path integral approach (QUAPI) \cite{MakriQUAPI1995a, MakriQUAPI1995b} which facilitates efficient numerical evaluation of the above path integral by reordering the influence functional and negelcting all bath memory influences beyond a given memory time $\tau_{\rm mem}$. Originally, QUAPI was developed for a system coupled to a single bath but it is easily extended to treat multiple baths if a common basis of all system-bath coupling operators, i.e. in our case $\sigma_1$ and $\sigma_2$, can be found \cite{NalbachNJP2010}.\\

\subsection{Non-Abelian Fluctuations}

When the various system-bath coupling operators do not commute, i.e. $0 \not = [\hat{\sigma_1},\hat{\sigma_2}]$, the influence functional is more evolved. We focus on a case with $[\hat{\sigma_1},\hat{\sigma_2}] \not = 0 = [\hat{\sigma_1},H_S]$ where $H_{SB,1}$ is purely a dephasing noise but $H_{SB,2}$ allows relaxation.

To derive the reduced density matrix components we first Trotter slice the propagator in $N$ time slices of length $\delta t_i=t/N$ and employ a symmetric Trotter splitting for the bath 2:
\bee
e^{-iH \delta t/\hbar} &\simeq & e^{-iH_{SB,2} \delta t/2\hbar} e^{-iH_S \delta t/\hbar}\cdot \\
&& \hspace*{1cm} \cdot  e^{-iH_{SB,1} \delta t\hbar} e^{-iH_{SB,2} \delta t/2\hbar} + O(\delta t^3) \nonumber
\eee
with error $O(\delta t^3)$ leading to $\rho_{S,r}(\sigma'_2 , \sigma''_2 ;t) = $
\begin{widetext}
\[ \label{rhored2bath2}
{\rm Tr}_B\left\{ \langle \sigma''_2 | e^{-i\frac{h_2}{2}} \left\{  \prod_{j=1}^{N-1} e^{-i(h_S+h_1)} e^{-i h_2} \right \} e^{-i(h_S+h_1)} e^{-i\frac{h_2}{2}}
\cdot \rho(0)  \cdot  
 e^{i\frac{h_2}{2}} e^{i(h_1+h_S)} \left\{  \prod_{j=1}^{N-1} e^{i h_2} e^{i (h_1+h_S)} \right \} e^{i\frac{h_2}{2}} | \sigma'_2 \rangle\right\}  .
\]
\end{widetext}
A more detailed derivation is given in appendix \ref{app3}. The symmetric splittings ensure that the total error of the time evolution are quadratic in $\delta t$, i.e. $O(N\delta t\cdot \delta t^2) = O(t\cdot \delta t^2)$. This finally allows to achieve convergence in the numerical treatment. We furthermore used the short hand notation 
\be\label{sh1}
h_{\alpha }=H_{SB,\alpha }\delta t/\hbar \quad{\rm and}\quad h_S=H_S\delta t/\hbar .
\ee
Inserting $2\cdot2\cdot N$ $\bbbone$ operators, i.e. $2N$ times $\bbbone_{1}$ and $2N$ times $\bbbone_2$ with
\bee
\bbbone_{1,j} &=& \int d\sigma_{1,j}^\pm \; |\sigma_{1,j}^\pm\rangle \langle \sigma_{1,j}^\pm | \nonumber \\
\bbbone_{2,j} &=& \int d\sigma_{2,j}^\pm \; |\sigma_{2,j}^\pm\rangle \langle \sigma_{2,j}^\pm | \nonumber
\eee
with $j\in\{0,1,\ldots,N-1\}$ as counting varibale leads to (see appendix \ref{app3} for details) $\rho_{S,r}(\sigma'_2 , \sigma''_2 ;t) =$
\begin{widetext}
\be
 \prod_{j=0}^{N-1} \int d\sigma_{1,j}^+ \int d\sigma_{1,j}^- \int d\sigma_{2,j}^+ \int d\sigma_{2,j}^-  \, K(\sigma^\pm_{1,j} ,\sigma^\pm_{2,j} , \sigma^\pm_{2,j+1} ) \cdot \langle \sigma_{2,0}^+ | \rho_S(0) |\sigma_{2,0}^-\rangle \cdot I(\{\sigma_{1,i}^\pm, \sigma_{2,i}^\pm : i=0\ldots N-1\}, \sigma_{2,N}^\pm)  \label{rhosfull}
\ee
\end{widetext}
with $\sigma_{2,N}^+=\sigma_2''$ and $\sigma_{2,N}^-=\sigma_2'$. 

Herein we definded the propagator
\bee  && K(\sigma^\pm_{1,j} ,\sigma^\pm_{2,j} , \sigma^\pm_{2,j+1}   )=
\langle \sigma_{2,j+1}^+ |\sigma_{1,j}^+ \rangle  
\langle \sigma_{1,j}^+ | e^{-iH_S \delta t/\hbar} |\sigma_{1,j}^+\rangle \nonumber\\
&& \hspace*{3mm}\times  
\langle \sigma_{1,j}^+ |\sigma_{2,j}^+ \rangle 
\langle \sigma_{2,j}^- |\sigma_{1,j}^- \rangle 
\langle \sigma_{1,j}^- | e^{iH_S \delta t/\hbar} |\sigma_{1,j}^- \rangle 
\langle \sigma_{1,j}^- |\sigma_{2,j+1}^- \rangle \nonumber
\eee
and the influence functional 
\bee
I(\{\sigma_{1j}^\pm, \sigma_{2,j}^\pm : j=0\ldots N-1\}, \sigma_{2,N}^\pm) &=& \nonumber \\
&& \hspace*{-5.2cm} I_1(\{\sigma_{1,j}^\pm : j=0\ldots N-1\})\cdot I_2(\{\sigma_{2,j}^\pm : j=0\ldots N\}) \nonumber
\eee
with
\bee
I_2(\{\sigma_{2,i}^\pm \}) 
&=& {\rm Tr}_{\rm B,2}\left\{ 
e^{-i \frac{h_2(\sigma_{2,N}^+)}{2}} \left\{ \prod_{j=1}^{N-1}
e^{-ih_2(\sigma_{2,j}^+) } \right\}  \right. \nonumber \\
&& \hspace*{-1.8cm} \left. e^{-i \frac{h_2(\sigma_{2,0}^+)}{2}}  \rho_{B,2}(0)  
e^{i\frac{h_2(\sigma_{2,0}^-)}{2}}  \left\{ \prod_{j=1}^{N-1}
e^{ih_2(\sigma_{2,j}^-) } \right\}   
e^{i \frac{h_2(\sigma_{2,N}^-)}{2}}  \right\} . \nonumber 
\eee
Therein, we used the shorthand notation $\{\sigma_{\nu ,j}^\pm \} = \{\sigma_{\nu,j}^\pm : j=0\ldots N_\nu\}$ with $N_1=N-1$ and $N_2=N$. Furthermore,
\[
h_\alpha |\sigma_\alpha\rangle = h_\alpha(\sigma_\alpha)|\sigma_\alpha\rangle
\]
and, accordingly, $h_\alpha(\sigma)$ is a bath operator acting solely on the Hilbert space of bath $\alpha$. Note that the influence functional $I_2(\{\sigma_{2,i}^\pm \})$ for bath $2$ is identical to the influence functional for a single bath as described by Makri and Makarov \cite{MakriQUAPI1995b}. 

The influence functional $I_1(\{\sigma_{1,j}^\pm\})$ of bath $1$ differs since it involves only $N-1$ time steps of length $\delta t$ thus missing the initial and final half-length steps in $I_2(\{\sigma_{2,j}^\pm \})$. In detail, we get
\bee
I_1(\{\sigma_{1,i}^\pm\}) &=& {\rm Tr}_{\rm B,1}\left\{ e^{-ih_1(\sigma_{1,N-1}^+) } \cdots  e^{-ih_1(\sigma_{1,0}^+) } \right. \nonumber \\
&& \hspace*{0cm} \left. \times 
\rho_{B,1}(0) \cdot 
e^{-ih_1(\sigma_{1,0}^+) } \cdots e^{-ih_1(\sigma_{1,N-1}^+) } 
 \right\} . \nonumber 
\eee

\subsection{Influence Functionals}

In thermal equilibrium to temperature $T$ the influence functional \cite{MakriQUAPI1995a, MakriQUAPI1995b} (in discrete form) of a single bath with bath spectral function $G(\omega)$ can be expressed as
\bee
I_F(\{x_j^\pm \};\{ \eta^{(F)}_{j,j'}\};\delta t,N) &=& \\
&&\hspace*{-3.8cm}\exp \left( -\sum_{j=0}^N \sum_{j'=0}^j (x_j^+-x_j^-) \left( \eta^{(F)}_{jj'}[\delta t] x_{j'}^+ - \eta^{(F) \star}_{jj'}[\delta t] x_{j'}^- \right) \right) \nonumber
\eee
for a time discretization of $N$ steps of size $\delta t$. The coefficients $\eta^{(F)}_{jj'}[\delta t]$ are explicitely given in appendix \ref{app2}.

As mentioned above, the influence functional for our bath 2 is identical to one of a single bath and we, thus, obtain immediately 
\[
I_2(\{\sigma_{2,j}^\pm \})=I_F(\{\sigma_{2,j}^\pm \};\{ \eta^{(2)}_{j,j'}\};\delta t,N)  
\]
where $\eta^{(2)}_{j,j'} = \eta^{(F)}_{j,j'}$ and the bath spectral function $G_2(\omega)$ is used in eq. (\ref{eqb1}).

The influence functional $I_1(\{\sigma_{1,j}^\pm\})$ of bath 1 differs from the aformentioned form leading to 
\[
I_1(\{\sigma_{1,j}^\pm \})=I_F(\{\sigma_{1,j}^\pm \};\{ \eta^{(1)}_{j,j'}\};\delta t,N-1) 
\]
with 
\bee
\eta^{(1)}_{j,j'}[\delta t] &=& \int_{-\infty}^\infty d\omega F(\omega) \cdot 4\sin^2\left( \omega \frac{\delta t}{2} \right) e^{-i\omega \delta t (j-j')} \nonumber\\ 
\eta^{(1)}_{jj}[\delta t] &=& \int_{-\infty}^\infty d\omega F(\omega) \cdot e^{-i\omega \delta t} \nonumber 
\eee
for all $0\le j\le N-1$, $0\le j'<j$ and, with $\beta=1/\kb T$,
\[
F(\omega) = \frac{G_1(\omega) \exp(\beta\hbar\omega/2)}{\omega^2 \sinh(\beta\hbar\omega/2)} .
\]

\section{QUAPI scheme}

Having now established an explicit representation for the time discretized influence functional of two non-Abelian fluctuation sources allows readily to implement a QUAPI scheme similar to the one introduced by Makri and Makarov \cite{MakriJMP1995, MakriQUAPI1995a, MakriQUAPI1995b}. The total influence functional depends now on $4(N+1)$ variables. QUAPI, however, restricts the influence functional to include memory only for $\Delta j_{\rm max}$ time steps. Then, a tensor product of the reduced density matrix for $\Delta j_{\rm max}$ time steps must be stored and popagated. For, this we need to store $n^{4(\Delta j_{\rm max}+1)}$ complex numbers for $n$ being the dimension of the system Hilbert space.  

Restricting the influence functionals to include (pairwise) correlations over maximally $\Delta j_{\rm max}$ time steps leads to
\be I_\nu(\{\sigma_{\nu,j}^\pm \}) \simeq \prod\limits_{\Delta j=0}^{\Delta j_{max}}\prod\limits_{j=0}^{N_\nu -\Delta j} I_{\nu,\Delta j}(\sigma^\pm_{\nu,j},\sigma^\pm_{\nu,j+\Delta j})
\ee
for $\nu=1,2$ and with $N_1=N-1$, $N_2=N$ and
\bee 
&& I_{\nu,\Delta j}(\sigma^\pm_{\nu,j},\sigma^\pm_{\nu,j+\Delta j}) = \nonumber \\
&& \exp\left[ -(\sigma^+_{\nu,j+\Delta j}-\sigma^-_{\nu,j+\Delta j})(\eta_{(j+\Delta j)j}^{(\nu)} \sigma^+_{\nu,j}-\eta_{(j+\Delta j)j}^{*,(\nu)} \sigma^-_{\nu,j}) \right] . \nonumber
\eee
Then, we define the propagator
\bee
&& \Lambda_j(\sigma^\pm_{1/2,j},...,      \sigma^\pm_{1/2,j+1+\Delta j_{max}} ) = K(\sigma^\pm_{1,j},\sigma^\pm_{2,j} ,\sigma^\pm_{2,j+1}  ) \nonumber \\               
&& \hspace*{1cm}  \times \prod_{\nu=1}^2 \left( I_{\nu,0}(\sigma^\pm_{\nu,j}) \dots I_{\nu,\Delta j_{max}}(\sigma^\pm_{\nu,j},\sigma^\pm_{\nu,j+\Delta j_{max}}) \right) \nonumber
\eee
using the  notation $\{\sigma^\pm_{1/2,j}\}=\{\sigma^\pm_{1,j},\sigma^\pm_{2,j}\}$ which allows to propagate the reduced density tensor 
\bee
&& A_{j+1}(\sigma^\pm_{1/2,j+1},..,\sigma^\pm_{1/2,j+\Delta j_{max}}) = \int d\sigma^\pm_{1,j}\int d\sigma^\pm_{2,j} \nonumber \\
&& \times A_j(\sigma^\pm_{1/2,j},..,\sigma^\pm_{1/2,j+\Delta j_{max}-1}) \cdot \Lambda_j(\sigma^\pm_{1/2,j},..,\sigma^\pm_{1/2,j+\Delta j_{max}}) \nonumber
\eee
with intial condition
\[ 
A_0(\sigma^\pm_{1/2,0},..  , \sigma^\pm_{1/2,\Delta j_{max}-1}  )=   \langle \sigma_{2,0}^+ | \rho_S(0) |\sigma_{2,0}^-\rangle . 
\]

The reduced density tensor $A_j(\cdot)$ is iteratively propagated. With a given $A_{j^\star}(\cdot)$ the reduced density matrix $\rho_{S,r}(\sigma'_2 , \sigma''_2 ;t)$ can be determined for a time $t=N\cdot \delta t$ with $N=j^\star + \Delta j_{\rm max}$ using
\bee
&& \rho_{S,r}(\sigma'_2 , \sigma''_2 ;t) = \int d\sigma^\pm_{1/2,{j^\star}} \dots \int d\sigma^\pm_{1/2,N-1} \nonumber \\
&& \hspace*{6mm}  \times A_{j^\star}(   \sigma^\pm_{1/2,{j^\star}}, ..,\sigma^\pm_{1/2,{j^\star}+\Delta j_{max}-1} ) \nonumber\\
&& \hspace*{6mm} \times K(\sigma^\pm_{1,{j^\star}} ,\sigma^\pm_{2,{j^\star}} , \sigma^\pm_{2,{j^\star}+1}   )...  K(\sigma^\pm_{1,N-1} ,\sigma^\pm_{2,N-1} , \sigma^\pm_{2,N}   ) \nonumber \\
&& \hspace*{6mm}  \times\prod\limits_{\Delta j=0}^{\Delta j_{max}}\prod\limits_{j={j^\star}}^{N-1-\Delta j} I_{1/2,\Delta j}(\sigma^\pm_{1/2,j},\sigma^\pm_{1/2,j+\Delta j})   \nonumber \\
&&  \hspace*{6mm} \times I_{2,0}(\sigma_{2,N})I_{2,1}(\sigma_{2,N-1},\sigma_{2,N})...I_{2,\delta j_{max}}(\sigma_{2,{j^\star}},\sigma_{2,N})  . \nonumber 
\eee
Note that here the influence functionals $\eta^{(2)}_{N,j}$ are employed which are not used within the iterative propagation of the A-tensor. This scheme follows the QUAPI scheme as described by Makri and Makarov \cite{MakriJMP1995, MakriQUAPI1995a, MakriQUAPI1995b} but extends it to a case with two non - Abelian noise sources. This scheme works for all times $t>\Delta j_{max}\cdot \delta t$. For the first $\Delta j_{max}$ time steps the reduced density matrix $\rho_{S,r}(\sigma'_2 , \sigma''_2 ;t)$ has to be calculated directly from Eq.(\ref{rhosfull}). Two explicit examples, i.e. $t=\delta t$ and $t=\Delta j_{max} \delta t$ are given in the appendix \ref{app1}. 

\begin{figure}[t]
\includegraphics[width=8.5cm]{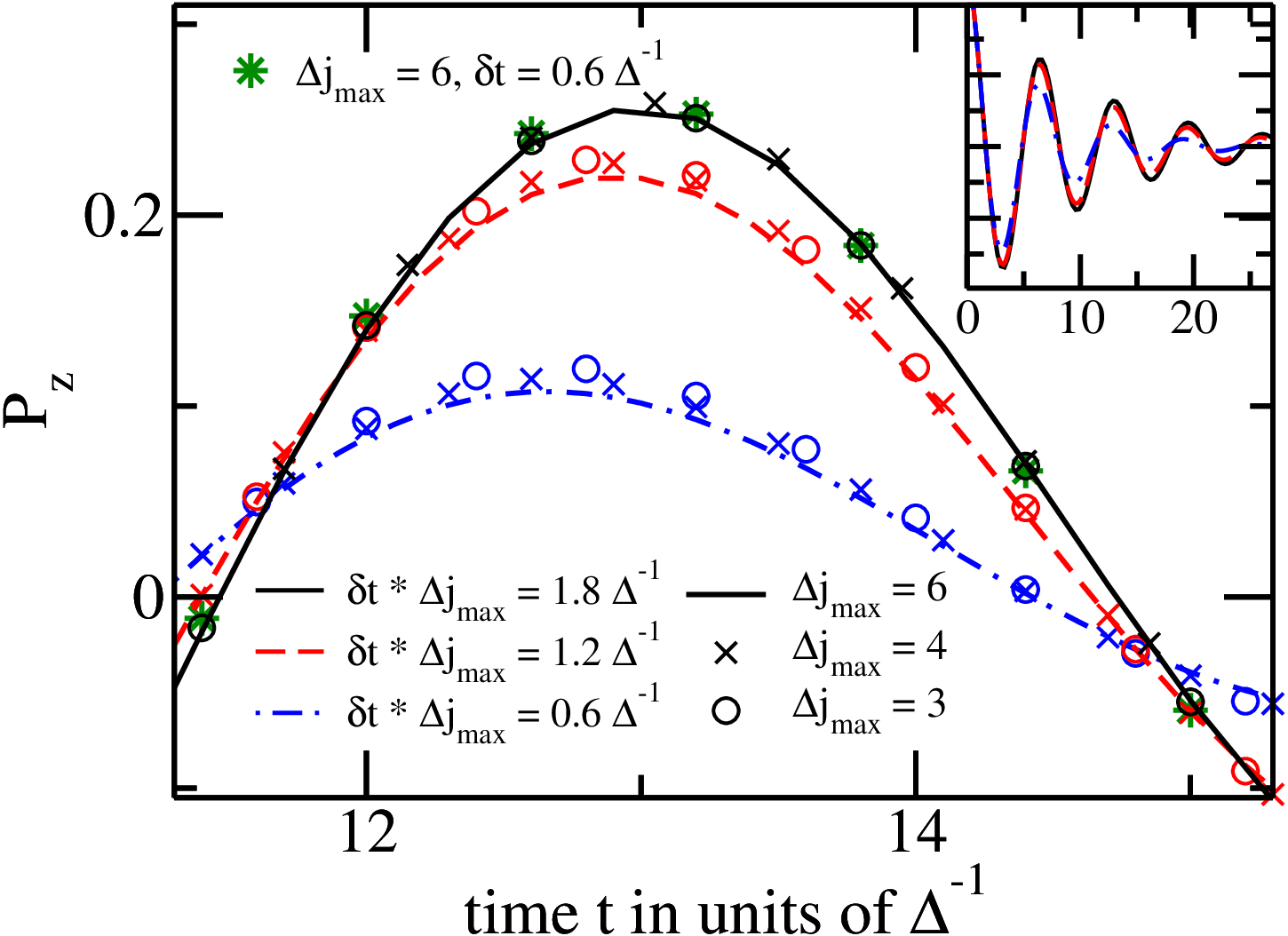}      
\caption{\label{fig1} (Color Online) Expectation value $P_z$ as function of time for a TLS disturbed by two environmental noise sources for various parameters $\delta t$ and $\Delta j_{\rm max}$.}
\end{figure}

\section{First Results and tests}

Exemplary we determine the dynamics of a quantum two-level system (TLS) with dipolar coupling $\Delta$ with Hamiltonian
\be\label{eq2.1}
H_S = \frac{\Delta}{2} \hat{\sigma}_x 
\ee
and Pauli matrices $\hat{\sigma}_\nu$. The TLS is under the influence of two independent non-commuting baths, i.e. bath 1 (in the following refered to as x-bath) couples via $\hat{\sigma}_x$ to the TLS and, thus, is a pure dephasing bath. Bath 2 (in the following refered to as z-bath) couples via $\hat{\sigma}_z$ which enables relaxation in the TLS.

Fig. \ref{fig1} shows the time evolution of $P_z(t)=\langle \hat{\sigma}_z\rangle(t)$ for various $\Delta j_{\rm max}$ and $\delta t$ where we have choosen for $P_z(0)=1$ and both bath to be in thermal equilibrium to temperature $T=0.2\Delta/k_{\rm B}$. We employed an Ohmic spectral density for both bath, i.e.
\be
G_\nu(\omega) = \frac{\gamma_\nu}{\pi} \omega e^{-\omega/\omega_c}
\ee
with cut-off frequency $\omega_c=10\Delta$ and coupling strengths $\gamma_z=1/16=\gamma_x$. The inset shows the studied time evolution for $\Delta j_{\rm max} = 6$ and three memory times, i.e. $\tau_{\rm mem} = \delta t \cdot \Delta j_{\rm max} = 1.8 \Delta^{-1}$, $1.2 \Delta^{-1}$ and $0.6 \Delta^{-1}$. The main figure shows the same time evolution for times restricted between $11 \Delta^{-1}$ and $15 \Delta^{-1}$. The data for $\Delta j_{\rm max} = 6$ and the three memory times are given by the full black, the dashed red and the dot-dashed blue line respectively. The black / red / blue circles (crosses) show data for the same memory time but $\Delta j_{\rm max} = 3$ ($\Delta j_{\rm max} = 4$). We see that different $\delta t$ (and, thus $\Delta j_{\rm max}$,) for identical memory times results in rather small deviations for $\tau_{\rm mem} = 1.2 \Delta^{-1}$ and $0.6 \Delta^{-1}$. For $\tau_{\rm mem} = 1.6 \Delta^{-1}$ the according differences are negligible. At the same time $P_z$ differs rather strongly for the three studied memory times. The largest memory time $\tau_{\rm mem} = 1.6 \Delta^{-1}$ reflects converged data since data with even larger memory time falls ontop of it as highlighted by data for $\Delta j_{\rm max} = 6$ and $\delta t = 0.6 \Delta^{-1}$ (green star symbols). In total, convergence can be found with identical approaches as used for the regular QUAPI code. 

\begin{figure}[t]
\includegraphics[width=8.5cm]{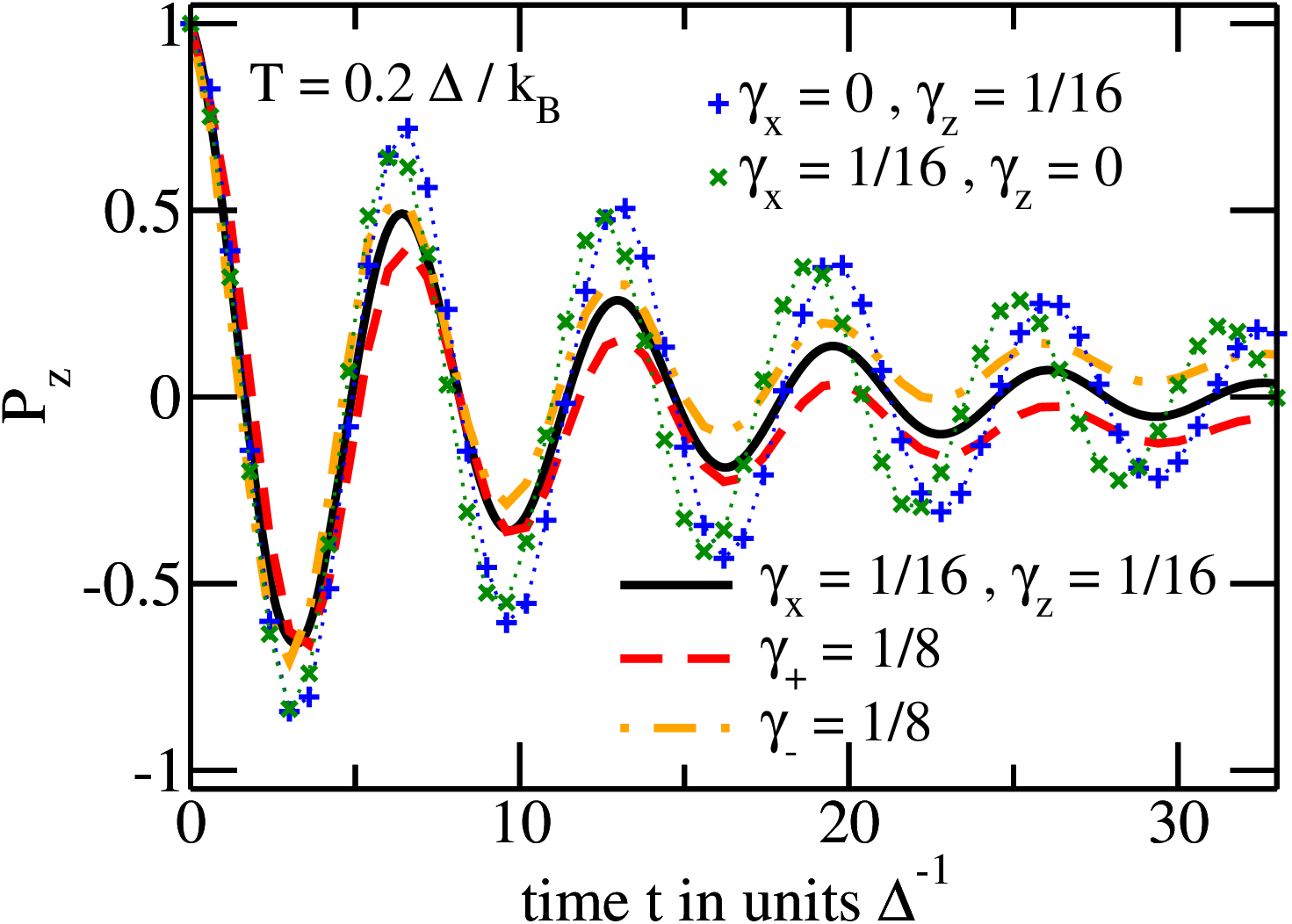}      
\caption{\label{fig2} (Color Online) Expectation value $P_z$ as function of time for a TLS disturbed by several environmental noise sources. Data shown for the two bath case is determined with $\delta t = 0.6 \Delta^{-1}$ and $\Delta j_{\rm max}=6$.}
\end{figure}

Converged results for the TLS under the influence of two separate bath (for the same parameters as before) are given by the full black line in Fig. \ref{fig2} which exhibits coherent oscilations which decay with a dephasing rate $\Gamma_{\rm deph}[\gamma_z=1/16;\gamma_x=1/16]=0.096\Delta$ towards $P_z\rightarrow 0$. The blue $+$ symbols in Fig.\ \ref{fig2} show $P_z$ for a TLS influenzed by the z-bath only, whereas the green x symbols reflect $P_z$ for a TLS influenzed by the x-bath only. Both exhibit coherent oscilations decaying towards zero with rates $\Gamma_{\rm deph}[\gamma_z=1/16;\gamma_x=0]=0.053\Delta$ and $\Gamma_{\rm deph}[\gamma_z=0;\gamma_x=1/16]=0.049\Delta$ respectively. Within a lowest order Redfield approximation \cite{May-Kuehn-2011} the dephasing rates for the z-bath is $\Gamma_{\rm deph,z} = \gamma_z\Delta\coth(\beta\Delta/2)= 0.063\Delta$. The discrepancy to the observed value of $\Gamma_{\rm deph}[\gamma_z=1/16;\gamma_x=0]=0.053\Delta$ is due to frequency renormalization of the not-so-weak system-bath coupling. The Redfield result for the x-bath is $\Gamma_{\rm deph,x} = 4 \gamma_z k_{\rm B} T = 0.05\Delta$ in close agreement with $\Gamma_{\rm deph}[\gamma_z=0;\gamma_x=1/16]=0.049\Delta$. In lowest order Redfield the dephasing rate for the TLS disturbed by both bath simultaneously is simply the sum $\Gamma_{\rm deph,x}+\Gamma_{\rm deph,z} = 0.102\Delta$ roughly 5\% larger than the observed $\Gamma_{\rm deph}[\gamma_z=1/16;\gamma_x=1/16]=0.096\Delta$ which again shows the onset of higher order effects due to not-so-weak system-bath coupling. We have also tested that for smaller system-bath couplings better agreement between our numerical results and weak coupling analytical estimates is achieved (data not shown).

Noise sources are typically difficult to analyze beyond their direct effect on a measurable system. Thus, when a system exhibits fluctuations in their $\langle\hat{\sigma}_z\rangle$ and $\langle\hat{\sigma}_x\rangle$ component, it might as well result from two independent bath as from a single bath coupling to both operators. In the latter case the fluctuations are fully correlated whereas in the former case they are uncorrelated. To model a single bath coupling to both operators, we need
\be
H_{SB,o_\pm} = \sum_{k=1}^M \frac{p_{k}^2}{2m_{k}} + \halb m_{k} \omega_{k}^2 \left( q_{k} - \frac{\lambda_{k} \hat{o}_\pm}{m_{k}\omega_{k}^2} \right)^2 
\ee
with $\hat{o}_\pm = \alpha_z \hat{\sigma}_z \pm \alpha_x\hat{\sigma}_x$ (in the following denoted as $o_\pm$-bath). In lowest order Redfield approximation this bath coupling exhibits as dephasing rate again the sum of the x-bath and the z-bath, i.e. $\Gamma_{\rm deph,x}+\Gamma_{\rm deph,z} = 0.102\Delta$ for both $\hat{o}_\pm$ if we choose $\alpha_{z/x} = \sqrt{\gamma_{z/x}/(\gamma_z+\gamma_x)}$ and the coupling strength in the according spectral density $\gamma_{\pm} = (\gamma_z+\gamma_x)$. The according time evolution for $P_z$ is given in Fig. \ref{fig2} by the red dashed line and the orange dot-dashed line. Again both exhibit coherent oscilations which decay with a dephasing rate $\Gamma_{\rm deph,\pm} = 0.109\Delta$. The dephasing rate for the two independent bath case was 10\% smaller which might be a first hint of frustrated decoherence as expected for very strong non-commuting system-bath couplings. 

The $o_\pm$-bath results via non-secular terms in lowest order Redfield to renormalizations of the eigenbasis of the system Hamiltonian. This causes shifted thermal equilibria resulting in $P_z\rightarrow \pm 0.08$ for $o_\pm$-bath. Thus, the steady state $P_z(t\rightarrow\infty)$ can be used to quantify the correlations in fluctuations of various expectation values, i.e. $\langle\hat{\sigma}_z\rangle$ and $\langle\hat{\sigma}_x\rangle$, and, thus, to determine whether a single noise source or mutliple independent noise sources are responsible for the fluctuations.

\begin{figure}[t]
\includegraphics[width=8.5cm]{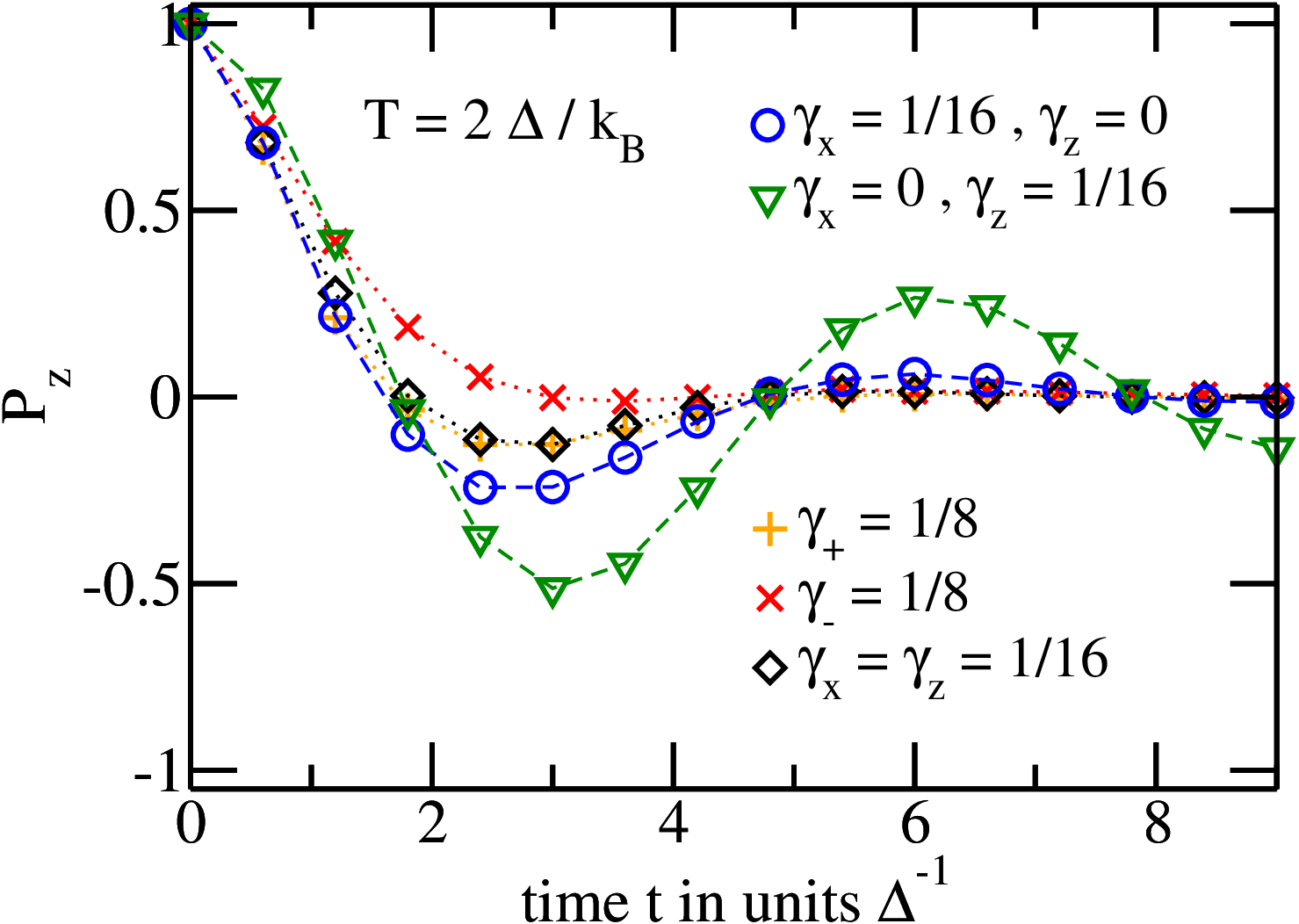}      
\caption{\label{fig3} (Color Online) Expectation value $P_z$ as function of time for a TLS disturbed by several environmental noise sources. Data shown for the two bath case is determined with $\delta t = 0.6 \Delta^{-1}$ and $\Delta j_{\rm max}=6$. }
\end{figure}

Finally, we should remark that only the z-bath causes a renormalization of the oscilating frequency in $P_z$. Accordingly, the x-bath case (green x symbols where $\gamma_z=0$) shows decaying oscilations $P_z\sim \cos(\Delta t)$. Surprisingly, all other cases have $P_z \sim \cos(\tilde{\Delta} t)$ with $\tilde{\Delta}\simeq 0.94\Delta$. Irrespective of whether the x-bath is present or not and of whether fluctuations are correlated or not ($o_\pm$-bath vs. two independent bath).

In Fig. \ref{fig3} we show $P_z$ for the same cases as before with identical parameters except that here we study an elavated temperature, i.e. $T=2\Delta/k_{\rm B}$. As expected, $P_z$ decays much faster. The dephasing rate in the x-bath case is roughly by a factor of 10 larger as expected due to the linear temperature dependence of the weak coupling Redfield rate $\Gamma_{\rm deph,x} = 4 \gamma_z k_{\rm B} T$. The z-bath dephasing rate is roughly larger by a factor of 4 due to $\coth(\Delta/(2* 2\Delta)) / \coth(\Delta/(2* 0.2\Delta)) \simeq 4$ which follows from the temperature dependence of the weak coupling Redfield rate $\Gamma_{\rm deph,z} = \gamma_z\Delta\coth(\beta\Delta/2)$. The dephasing rates for the $o_\pm$-baths as well as for the two independent bath cases are roughly the sum of the two former cases where the rate for the two independent bath case is roughly 3\% smaller. The steady state for all cases is $P_z\rightarrow 0$. Thus, increased temperature erases the correlation effects in the fluctuations.

\section{Conclusions}

Typical quantum systems are subject to various environmental noise sources. If one of them is strongly influencing the quantum system, the others cannot be treated within weak coupling approaches either \cite{PalmPRA2017}. We have extended the numerically exact quasi-adiabatic path integral approach \cite{MakriQUAPI1995a, MakriQUAPI1995b} which allows to determine the time dependent reduced statistical operator of a quantum system under the influence of multiple noise sources. For baths which couple to the system via operators, which commute, an according extension of QUAPI is available \cite{NalbachNJP2010}. For the non-commuting case we now determined the necessary time discrete form of the total influence functional and extended the QUAPI scheme accordingly.

We test the extended QUAPI by determining the time evolution of a quantum two-level system coupled to two independent bath via non-commuting operators, i.e. $\hat{\sigma}_z$ and $\hat{\sigma}_x$. We find converged results following a scheme which increases the memory time $\tau_{\rm mem}=\Delta j_{\rm max}\cdot \delta t$ with decreasing the Trotter time slicing $\delta t$ at the same time. Convergence is reached when the results stay the same when decreasing $\delta t$ and increasing $\tau_{\rm mem}$ further. We then compare converged results with various single bath cases and observe results identical to standard analytical approximative results for weak coupling of all bath. In the future we can now employ our method to study the peculiar non-perturbative effects like quantum frustration of decoherence \cite{CastroPRL2003, KehreinPRB2013} due to strong non-commuting fluctuations at finite temperatures and their relevance in energy transfer in photosynthetic complexes and dephasing in various qubit realizations.

%
TP and PN acknowledge financial support by the DFG project NA394/2-1.

\appendix

\section{Time slicing of the reduced density matrix}\label{app3}

To derive the reduced density matrix components one Trotter-slices the propagator in $N$ time slices of length $\delta t_i=t/N$ and employ a symmetric Trotter splitting for the bath 2:
\bee
e^{-iH \delta t/\hbar} &\simeq & e^{-iH_{SB,2} \delta t/2\hbar} e^{-iH_S \delta t/\hbar}\cdot \nonumber \\
&& \hspace*{1cm} \cdot  e^{-iH_{SB,1} \delta t\hbar} e^{-iH_{SB,2} \delta t/2\hbar} + O(\delta t^3) \nonumber
\eee
with error $O(\delta t^3)$ leading to
\begin{widetext}
\bee \label{rhored2bath2b}
\rho_{S,r}(\sigma'_2 , \sigma''_2 ;t) &=& {\rm Tr}_B\left\{ \langle \sigma''_2 | e^{-iHt/\hbar} \rho(0) e^{iHt/\hbar} | \sigma'_2 \rangle\right\}  = {\rm Tr}_B\left\{ \langle \sigma''_2 | \prod_{j=0}^{N-1} e^{-iH\delta t_j/\hbar} \cdot \rho(0) \cdot  \prod_{j=0}^{N-1} e^{iH \delta t_j/\hbar} | \sigma'_2 \rangle\right\} \nonumber \\
&\hspace*{-4cm}=& \hspace*{-2cm} {\rm Tr}_B\left\{ \langle \sigma''_2 | e^{-iH_{SB,2}\delta t_{N-1}/2\hbar} \left\{ \prod_{j=1}^{N-1} e^{-iH_S\delta t_j/\hbar} e^{-iH_{SB,1} \delta t_j/\hbar} e^{-iH_{SB,2}\frac{(\delta t_{j}+\delta t_{j-1})}{2\hbar}} \right\} \cdot e^{-iH_S\delta t_0/\hbar} e^{-iH_{SB,1} \delta t_0/\hbar} e^{-iH_{SB,2}\delta t_{0}/2\hbar} \right. \nonumber\\
&& \left. \hspace*{-1cm} 
\cdot \rho(0)  \cdot  e^{iH_{SB,2}\delta t_{0}/2\hbar} e^{iH_S\delta t_0/\hbar} e^{iH_{SB,1} \delta t_0/\hbar} \left\{ \prod_{j=1}^{N-1}  e^{iH_{SB,2}\frac{(\delta t_{j}+\delta t_{j-1})}{2\hbar}} e^{iH_{SB,1} \delta t_j/\hbar} e^{iH_S\delta t_j/\hbar} \right\} \cdot e^{iH_{SB,2}\delta t_{N-1}/2\hbar} | \sigma'_2 \rangle\right\} \nonumber \\
&\hspace*{-4cm}=& \hspace*{-2cm} {\rm Tr}_B\left\{ \langle \sigma''_2 | e^{-i\frac{h_2}{2}} \left\{  \prod_{j=1}^{N-1} e^{-ih_S} e^{-ih_1} e^{-i h_2} \right \} e^{-ih_S} e^{-i h_1} e^{-i\frac{h_2}{2}}
\cdot \rho(0)  \cdot  
 e^{i\frac{h_2}{2}}e^{i h_1} e^{ih_S} \left\{  \prod_{j=1}^{N-1} e^{i h_2} e^{i h_1} e^{ih_S} \right \} e^{i\frac{h_2}{2}} | \sigma'_2 \rangle\right\} \nonumber \\
&\hspace*{-4cm}=& \hspace*{-2cm} {\rm Tr}_B\left\{ \langle \sigma''_2 | e^{-i\frac{h_2}{2}} \left\{  \prod_{j=1}^{N-1} e^{-ih_S} e^{-ih_1}\bbbone_{1,j} e^{-i h_2}\bbbone_{2,j} \right \} e^{-ih_S} e^{-i h_1}\bbbone_{1,0} e^{-i\frac{h_2}{2}}\bbbone_{2,0}
\cdot \rho(0) \right. \nonumber \\
&& \left. \hspace*{6.2cm}  \times  
 \bbbone_{2,0} e^{i\frac{h_2}{2}} \bbbone_{1,0} e^{i h_1} e^{ih_S} \left\{  \prod_{j=1}^{N-1} \bbbone_{2,j} e^{i h_2} \bbbone_{1,j} e^{i h_1} e^{ih_S} \right \}  e^{i\frac{h_2}{2}} | \sigma'_2 \rangle\right\} \nonumber \\
&\hspace*{-4.4cm}=& \hspace*{-2.2cm}  \prod_{j=0}^{N-1} \int d\sigma_{1,j}^+ \int d\sigma_{1,j}^- \int d\sigma_{2,j}^+ \int d\sigma_{2,j}^-  \, K(\sigma^\pm_{1,j} ,\sigma^\pm_{2,j} , \sigma^\pm_{2,j+1} ) \cdot \langle \sigma_{2,0}^+ | \rho_S(0) |\sigma_{2,0}^-\rangle \cdot I(\{\sigma_{1,i}^\pm, \sigma_{2,i}^\pm : i=0\ldots N-1\}, \sigma_{2,N}^\pm)  \nonumber 
\eee
\end{widetext}
with $\sigma_{2,N}^+=\sigma_2''$ and $\sigma_{2,N}^-=\sigma_2'$. 
In the second last line we have inserted the $\bbbone$ - operator to obtain the final time sliced reduced density matrix. 
The symmetric Trotter splittings ensure that the total error of the time evolution are quadratic in $\delta t$, i.e. $O(N\delta t\cdot \delta t^2) = O(t\cdot \delta t^2)$. This finally allows to achieve convergence in the numerical treatment. We furthermore used the short hand notation 
\[\label{sh1b}
h_{\alpha }=H_{SB,\alpha }\delta t/\hbar \quad{\rm and}\quad h_S=H_S\delta t/\hbar .
\]

\section{The $\eta$ - coefficients for a single bath}\label{app2}

The coefficients \cite{MakriQUAPI1995a, MakriQUAPI1995b} $\eta^{(F)}_{jj'}[\delta t]$ which express the bath correlations within the influence functional of a single bath with bath spectral function $G(\omega)$ at temperature $T$ are 
\bee
\eta^{(F)}_{jj'}[\delta t] &=& \int_{-\infty}^\infty d\omega F(\omega) \cdot 4\sin^2\left( \omega \frac{\delta t}{2} \right) e^{-i\omega \delta t (j-j')} \nonumber\\
\eta^{(F)}_{jj}[\delta t] &=& \int_{-\infty}^\infty d\omega F(\omega) \cdot e^{-i\omega \delta t} \nonumber\\
\eta^{(F)}_{N0}[\delta t] &=& \int_{-\infty}^\infty d\omega F(\omega) \cdot 4\sin^2\left( \omega \frac{\delta t}{4} \right) e^{-i\omega (t- \delta t/2)} \nonumber\\
\eta^{(F)}_{00}[\delta t] &=& \eta_{NN}[\delta t] = \eta_{N0}[\delta t/2] \nonumber\\
\eta^{(F)}_{j0}[\delta t] &=& \nonumber \\
&& \hspace*{-0.8cm} \int_{-\infty}^\infty d\omega F(\omega) \cdot 4\sin\left( \omega \frac{\delta t}{4} \right) \sin\left( \omega \frac{\delta t}{2} \right) e^{-i\omega (j\delta t- \delta t/4)} \nonumber\\
\eta^{(F)}_{Nj}[\delta t] &=& \nonumber \\
&& \hspace*{-1cm} \int_{-\infty}^\infty d\omega F(\omega) \cdot 4\sin\left( \omega \frac{\delta t}{4} \right) \sin\left( \omega \frac{\delta t}{2} \right) e^{-i\omega (t-j\delta t- \delta t/4)} \nonumber
\eee
for $0<j'<j<N$, with $\beta=1/\kb T$ and
\be\label{eqb1}
F(\omega) = \frac{G(\omega) \exp(\beta\hbar\omega/2)}{\omega^2 \sinh(\beta\hbar\omega/2)} .
\ee

\section{Explicit short time propagation} \label{app1}

For the first $\Delta j_{max}$ time steps the reduced density matrix $\rho_{S,r}(\sigma'_2 , \sigma''_2 ;t)$ has to be calculated directly from Eq.(\ref{rhosfull}). The first step, i.e. $t=\delta t$ is
\bee
\rho_{S,r}(\sigma'_2 , \sigma''_2 ;\delta t) &=& \int d\sigma^\pm_{1/2,0}K(\sigma^\pm_{1,0} ,\sigma^\pm_{2,0} , \sigma^\pm_{2,1}   )  \nonumber\\ 
&& \hspace*{-2.8cm}\cdot \langle \sigma_{2,0}^+ | \rho_S(0) |\sigma_{2,0}^-\rangle  I_{0,1}(\sigma^\pm_{1,0})I_{0,2}(\sigma^\pm_{2,0})I_{0,2}(\sigma^\pm_{2,1})I_{2,1}(\sigma^\pm_{2,0},\sigma^\pm_{2,1}) \nonumber
\eee
where $I_{2,1}$ includes $\eta^{(2)}_{N,0}$ due to $t_N=\delta t$.

For $t=\Delta j_{max} \delta t$ we obtain
\bee
\rho_{S,r}(\sigma'_2 , \sigma''_2 ;\Delta j_{max} \delta t) &=& \int d\sigma^\pm_{1/2,0} \dots \int d\sigma^\pm_{1/2,\Delta j_{max}-1}   \nonumber\\ 
&& \hspace*{-3.2cm}\cdot K(\sigma^\pm_{1,0} ,\sigma^\pm_{2,0} , \sigma^\pm_{2,1}   ) \dots K(\sigma^\pm_{1,N-1} ,\sigma^\pm_{2,N-1} , \sigma^\pm_{2,N}   ) \nonumber \\
&& \hspace*{-3.2cm}    \cdot \langle \sigma_{2,0}^+ | \rho_S(0) |\sigma_{2,0}^-\rangle \prod\limits_{\Delta j=0}^{\Delta j_{max}}\prod\limits_{j'=0}^{N-1-\Delta j} \prod_{\nu=1}^2 I_{\nu,\Delta j}(\sigma^\pm_{\nu,j},\sigma^\pm_{\nu,j+\Delta j})        \nonumber \\
&& \hspace*{-3.2cm} \cdot I_{2,0}(\sigma_{2,N})I_{2,1}(\sigma_{2,N-1},\sigma_{2,N})...I_{2,N}(\sigma_{2,0},\sigma_{2,N}) \nonumber
\eee
where $I_{2,N}$ uses $\eta_{N,0}$ due to $t_N=N\delta t$.


\end{document}